# Adaptive Privacy-Preserving SSD

Na Young Ahn

School of Cybersecurity at Korea University, humble@korea.ac.kr

Dong Hoon Lee

School of Cybersecurity at Korea University, donghlee@korea.ac.kr

Data remanence in NAND flash complicates complete deletion on IoT SSDs. We design an adaptive architecture offering four privacy levels (PL0-PL3) that select among address, data, and parity deletion techniques. Quantitative analysis balances efficacy, latency, endurance, and cost. Machine-learning adjusts levels contextually, boosting privacy with negligible performance overhead and complexity.

## 1 INTRODUCTION

By harnessing diverse storage paradigms—cloud, edge, blockchain, and on-device—to revolutionize daily life, the IoT also incurs serious privacy risks from pervasive data collection and sharing; this study proposes practical security measures to mitigate those risks while maintaining efficiency [1]. Privacy risks in IoT threaten individuals and society by undermining trust, incurring financial, reputational, and legal damage from breaches, and are exacerbated by NAND-flash data remanence—which resists full erasure, complicates forensic tracing, and lets attackers exploit residual fragments—making robust, balanced protection strategies indispensable. In IoT environments, it is imperative to address the data remanence problem associated with NAND flash memory meticulously [2-5,14]. Robust security measures, including secure data deletion, encryption, and personal data protection, are essential. These measures serve to safeguard personal privacy while aiding forensic analysts and law enforcement agencies in effectively utilizing data while adhering to legal and ethical principles [6].

This paper presents an adaptive SSD design to address these challenges. Our **Adaptive Privacy-Preserving SSD** dynamically adjusts its deletion strategy according to user-defined privacy levels and system constraints. We make the following key contributions:

- **Adaptive Privacy Levels:** We define four increasing privacy levels (PL0–PL3) with corresponding deletion methods: full block erase (PL0), targeted page scrubbing/overwrite (PL1), parity/ECC-based erasure (PL2), and controller-level block disablement (PL3). This enables tuning of privacy vs. performance.
- **Deletion Method Classification:** We categorize secure deletion approaches into three classes – address management, data erasure, and parity/ECC managementfile-aqbd1yjbourpdqkdsbs3vm – and systematically analyze their effectiveness, speed, endurance impact, verification capability, and cost.
- **AI-Aware Privacy Control:** We outline how integrating context-aware algorithms (e.g. ML-based sensitivity assessment) can automate the choice of privacy level in real time.
- **Cost-Centric Framework:** We evaluate each method on practical metrics (Table 1) and illustrate how costs and benefits change across privacy levels. This provides a guideline for designers to choose appropriate deletion schemes under IoT constraints.

This study investigates IoT privacy protection, focusing on secure deletion for 3D-NAND-based SSDs and offering guidance that reconciles data security with sustainable IoT growth. Section 2 examines 3D NAND storage architecture, Section 3 analyzes block-erase costs, and Sections 4–5 introduce and evaluate cost-aware, privacy-level-adaptive deletion techniques. By framing privacy levels as a cost metric, the proposed adaptive SSD approach establishes a practical benchmark for deployable privacy-preserving storage in the emerging IoT landscape.

## 2 BACKGROUND: 3D NAND FLASH MEMORY AND DATA MANAGEMENT CHALLENGES

3D NAND flash stacks hundreds of vertical layers linked by channels to provide far greater density, capacity, speed, and endurance than planar NAND, yet deletions still occur at block granularity, so updating a single page creates write amplification that inflates latency, power, wear, and ultimately cost. To curb this effect, the array carves out spare pages for mapping tables, ECC, and over-provisioning, making meticulous metadata management—aimed at minimizing block erases—vital for preserving SSD reliability and lifespan. Micron has previously announced the production of NAND flash memory with a 232-layer structure [7]. Micron's 232-layer 3D NAND surpasses the 176-layer design, offering greater capacity, speed, reliability, and endurance for high-density drives and cloud data centers while the company keeps advancing performance. However, whole-block erase semantics mean higher bits-per-cell inflate block counts and trigger more frequent, power-hungry erase–backup–restore cycles that harm performance unless mitigated by block grouping and aggressive garbage collection.

### 2.1 Increase in block erase cost

Generally, memory blocks are guaranteed for a limited number of erase cycles, typically less than a few thousand times. This limitation implies an inherent inability to guarantee the data in memory blocks permanently due to variations in charge leakage rates. Therefore, techniques that minimize the erase of memory blocks are widely employed. These management techniques inevitably raise considerations regarding associated costs [8]. Erasing at the block level forces every page in the block to be deleted and rewritten, consuming extra time, power, and controller resources, which slows writes, accelerates transistor wear, shortens drive life, and inflates garbage-collection overhead. Minimizing erase frequency through smarter block utilization and more efficient garbage collection is therefore critical to curbing these costs and preserving SSD performance.

### 2.2 Trad-off between block erase cost and privacy exposure

There is a direct relationship between increased block erase costs and the risk of personal data exposure. As block erase costs rise and available blocks diminish, if blocks aren't used efficiently, there may be no space left for data storage. When an SSD allocates new blocks for data storage, any residual data in the old blocks may become exposed. Residual data refers to data that has been logically deleted by the host but remains physically undeleted on the SSD device. If the logical mapping relationship is restored, this data can be fully recovered and reused. Such residual data has been observed to persist for several months or more in IoT devices [5]. Initializing whole blocks during deletion can leave residual fragments, so page-level sanitization, thorough verification, and post-erase overwriting are essential to eliminate exposure. Beyond privacy, reducing block-erase frequency bolsters SSD performance, longevity, and energy efficiency, motivating manufacturers to prioritize low-cost erase-management strategies.

### 2.3 Limitations of existing encryption technology

Even when personal information is encrypted, if the data is not completely erased and remains within a block, it is possible to decrypt that data, exposing the personal information. This risk arises because if encrypted data remains within a block, there is a chance that the block will be reused, making it feasible to decrypt and reveal personal details. In such circumstances, it's essential to completely erase encrypted data. This necessitates performing a block erase operation that initializes all pages within the block. If only a subset of the pages within the block is initialized without a block erase, encrypted data could persist, leading to the possibility of decryption and subsequent exposure of personal information [6]. This potential for exposure from encrypted data arises due to two primary factors: The presence of the encrypted data within the block. The potential for only partial page initialization during block erase operations. Also, the forensic technique of reading residual data from NAND flash memory is one of the crucial technologies in the field of digital forensics. This technique is employed to recover and analyze previously stored information from NAND flash memory devices, even when data has been completely deleted.

### 2.4 Access mechanism of unmapped blocks

Unmapped blocks in NAND flash create forensic vulnerabilities, because administrators—or attackers with similar privileges—can access these areas and potentially recover residual data. Although garbage-collection routines normally reinitialize such blocks, incomplete cycles can leave fragments that sophisticated scanning may extract. Block-level erasure further amplifies writes, forcing entire blocks to be rewritten for single-page updates, accelerating wear and shortening SSD lifespan. Spare pages that hold mapping tables and ECC bits mitigate errors but, in high-density 3D NAND, also raise erase costs, demanding careful wear-leveling and partial-programming strategies. Consequently, effective secure-deletion designs must confront unmapped-block exposure, write-amplification overhead, endurance limits, and metadata consistency to protect privacy without compromising performance.

## 3 PRIVACY RISKS AND SECURE DELETION CHALLENGES IN SSDs

Personal data stored on SSDs faces persistent privacy risks due to residual data fragments that survive logical deletion processes. Data duplication methods, while improving reliability, can increase data persistence and complicate simultaneous secure deletion across all copies. Encryption alone is insufficient, as encrypted data remnants may be decrypted or exposed, especially with the advent of quantum computing capabilities. Furthermore, unmapped or over-provisioned blocks, not tracked by the file system, can harbor residual or malicious data, posing additional security concerns. These challenges underscore the necessity for comprehensive secure deletion mechanisms that address both logical and physical data remnants in NAND flash memory. As mentioned, data remanence issues in unmapped blocks of NAND flash memory are inherent, but eradicating lingering data directly from NAND flash memory in terms of management is not a straightforward task. It is essential to recognize this as a management factor and consider the associated costs of achieving complete deletion.

### 3.1 Threats of commercialized quantum computing

Quantum computing is a technology capable of processing data at speeds much faster than traditional digital computing. This advantage can facilitate the decryption of encrypted data, increasing the likelihood of privacy exposure [10]. Quantum computing's accelerated decryption capabilities threaten conventional encryption—especially RSA and ECC and, eventually, even large-key symmetric ciphers like AES—so encrypted SSD data may become readable despite remaining "encrypted." Because quantum attacks also hamper key exchange and complicate secure deletion by enabling analysis of residual bits, storage security must incorporate quantum-resistant cryptography alongside more rigorous sanitization strategies. Consequently, the commercialization of quantum hardware demands simultaneous advances in key management, data-eradication techniques, and quantum-safe algorithms to avert persistent privacy exposure.

### 3.2 Threats due to data duplication

Data duplication boosts reliability by keeping copies in multiple locations, but it also magnifies privacy risk because every replica must be deleted synchronously. Residual copies persist when even one storage node fails to erase data, creating a clear data-persistence vulnerability. Uniform safeguards—encryption, access control, and auditing—must be applied consistently across all duplicated sites. If a single location lacks adequate protection, sensitive information deleted elsewhere remains recoverable from that unsecured replica. Consequently, duplication strategies require coordinated, system-wide deletion protocols and robust, uniform security measures.

### 3.3 Threats in unmapped blocks

Unmapped blocks in the file system are sections that aren't in use but can store data. While it's possible to store malicious code in these blocks, it's generally unlikely. Usually, for malicious code to infiltrate and run in a system, it must be stored as an executable file. However, unmapped blocks aren't registered in the file system, preventing them from holding executable files. Nevertheless, if there exists malicious software capable of executing code internally within an SSD, it might store malicious code in unmapped blocks and run it. This could occur if vulnerabilities in SSD firmware or controllers are exploited [8]. Although running malicious code from unmapped or over-provisioned SSD blocks is difficult, determined attackers can exploit these regions via firmware weaknesses. Users should counter this threat by selecting reputable, security-certified drives, keeping firmware patched, and restricting over-provisioning adjustments so excess space cannot hide rogue code. Consistent updates and carefully sized spare areas thus close a key avenue for low-level compromise.

## 4 PROPOSED ADAPTIVE PRIVACY-PRESERVING SSD ARCHITECTURE

We propose an adaptive privacy-preserving SSD design that dynamically adjusts secure deletion methods according to privacy requirements and operational cost constraints. Four privacy levels (PL0 to PL3) correspond to increasingly stringent deletion techniques: from full block erase (PL0), page-level scrubbing and overwriting (PL1), parity-level ECC disruption and read locking (PL2), to physical map-out of NAND blocks (PL3). This graduated approach enables users or automated systems to balance privacy assurance against performance overhead and device wear. Moreover, integrating artificial intelligence allows real-time assessment of data sensitivity and environmental factors, facilitating automatic privacy level adjustment to optimize security and efficiency in diverse IoT scenarios.

**4.1 Emergence of SSDs supporting adaptive privacy**

NAND flash memory is organized into pages and blocks, and rather than overwriting data, it can erase and write anew. The erase operation is executed at the block level, incurring certain costs. This 'cost' can encompass physical wear, time, and energy. While memory might appear entirely erased, residual data can remain. This leftover data potentially provides attackers access to personal or critical information. Consequently, securely erasing memory might necessitate multiple erase cycles or specialized algorithms, which can lead to additional costs.

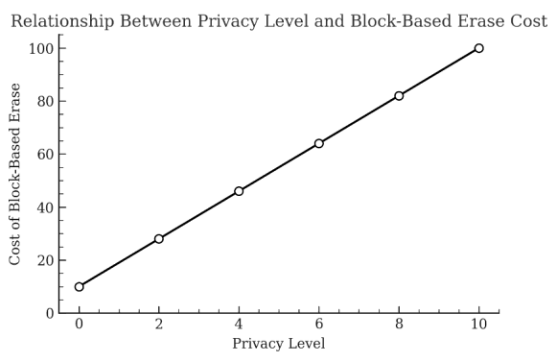

Figure 1: This illustratively demonstrates the relationship between privacy levels and block erase costs.

Referring to Fig. 1, it is evident that desiring a higher privacy level can escalate block erase costs. For instance, erasing a memory block might demand more time, energy, and operations. Flash memory's durability is finite. Therefore, erasing blocks too frequently can lead to a reduced memory lifespan. Adaptive-privacy-level SSDs let users or onboard sensors dynamically raise or lower protection, balancing data importance and environmental risk. Because stronger measures like secure erase or encryption demand more time, energy, and space, privacy and performance are inherently traded off—a relationship illustrated in our cost graph. The device can thus run minimal safeguards on trusted home networks yet invoke intensive deletion or encryption in hostile settings or for critical files such as financial records. Users retain manual control over these levels, while context-aware firmware can auto-adjust them by monitoring network type, location, or threat cues. Implementing such flexibility requires tight integration of encryption engines, policy management, user interfaces, and environmental sensors, delivering both usability and security across diverse scenarios.

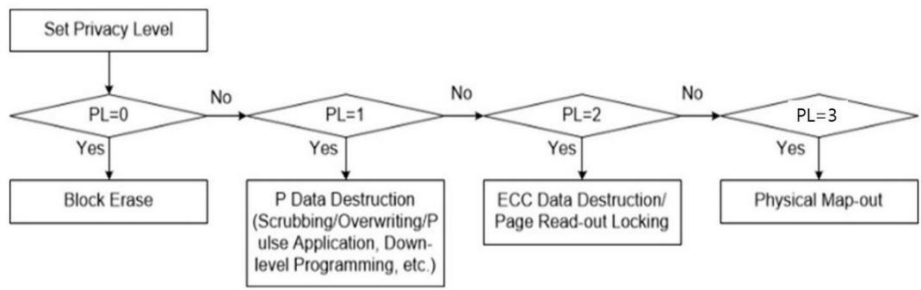

Figure 2: This conceptually illustrates the operation of the proposed storage device supporting Adaptive Privacy Levels.

For the sake of clarity, Fig. 2 illustrates only four privacy levels (PL0 ~ PL3). Based on the set privacy levels (PL0 ~ PL3), secure deletion can proceed as follows:

Level 1 (PL=0): Implements secure deletion using block erase.

Level 2 (PL=1): Performs secure deletion by destroying privacy data at the page level. At this stage, existing secure deletion techniques such as scrubbing [12], overwriting [12], deletion pulse application [13], or down-level programming [13] can be utilized.

Level 3 (PL=2): Executes secure deletion using error correction code destruction or page read-out locking. Both the destruction of error correction codes and page read-out locking are techniques that address secure deletion in the spare region of a page.

Level 4 (PL=3): Implements map-out on physically invalid areas, thereby permanently blocking access to the corresponding area of the NAND flash memory from the controller.

When the privacy level is set at the start, the system sequentially evaluates each privacy level condition from PL=0 up to PL=3, as depicted by the series of decision diamonds. If the level is **PL=0**, the **Block Erase** operation is performed, wiping the entire block of memory. If the privacy level is not 0, the flow proceeds to check **PL=1**: a "Yes" at PL=1 triggers *P Data Destruction* – this involves targeted data sanitization methods such as scrubbing, overwriting, applying voltage pulses, or down-level programming to erase sensitive data. Similarly, a match at **PL=2** initiates **ECC data destruction** or **page read-out locking**, leveraging error-correcting code mechanisms to corrupt or lock away the data. Finally, if the set level is **PL=3** (the highest privacy level), a **Physical Map-out** is executed, physically isolating or disabling the memory cells (e.g., marking them as bad blocks) to prevent any future access to the data. In this adaptive scheme, each increment in privacy level corresponds to a more intensive secure erase method, ensuring that the data sanitization rigor matches the required privacy protection level.

Table 1: Privacy Level Comparison (PL0–PL3)

| Privacy Level | Deletion Method(s) | Latency | Endurance Impact | Forensic Resistance | Implementation Feasibility |
|---|---|---|---|---|---|
| PL0 | Block Erase | High | High | High | High |
| PL1 | Page-level (Scrubbing, Overwrite, Deletion Pulses, Down-level Prog.) | Medium | Medium | Medium | Medium |
| PL2 | ECC Parity Destruction (Read-Out Locking) | Low | Low | Medium | Medium |
| PL3 | Physical Map-Out of Block | Low | Low | High | Low |

Table 1 compares privacy levels PL0–PL3, their deletion methods, and key trade-offs. PL0 relies on a full block erase that physically removes all data, delivering high forensic resistance. This strategy suffers high latency because entire blocks must be reset. It also imposes heavy wear, reducing SSD endurance. PL1 moves to page-level techniques—scrubbing, overwriting, erase pulses, or down-level programming—which trims latency to a medium level. Page-level deletion still perturbs neighboring cells, so its endurance impact remains moderate. PL2 uses parity-management tactics such as ECC-parity destruction or read-out locking, providing low latency and minimal extra wear. By forcing ECC failure, PL2 renders the page unreadable and offers medium-to-high anti-forensic strength even though raw bits persist. PL3 attains the highest logical privacy by invalidating the physical region, denying controller access with negligible latency or wear. Although normal-access forensic resistance is high, PL3's permanent bad-block mapping raises implementation complexity and sacrifices usable capacity.

Our adaptive privacy-preserving SSD can embed AI that dynamically sets privacy levels, shielding sensitive data from unauthorized access while simplifying user management. By evaluating data sensitivity and context in real time, the AI elevates protection for critical information, relaxes it for benign content, and automatically removes superfluous files, optimizing capacity and performance. This intelligent gradation—from basic erasure to sophisticated access denial—delivers strong security, streamlined usability, and efficient storage management.

**4.2 Various schemes of Secure Deletion**

Existing secure deletion techniques for invalid data areas in storage devices can be categorized into: 1) Address Management Techniques [11], 2) Data Management Techniques [13], and 3) Parity Management Techniques [12,15]. This classification provides a systematic understanding of how data security is maintained across different layers of storage management, ensuring the robust protection of sensitive information.

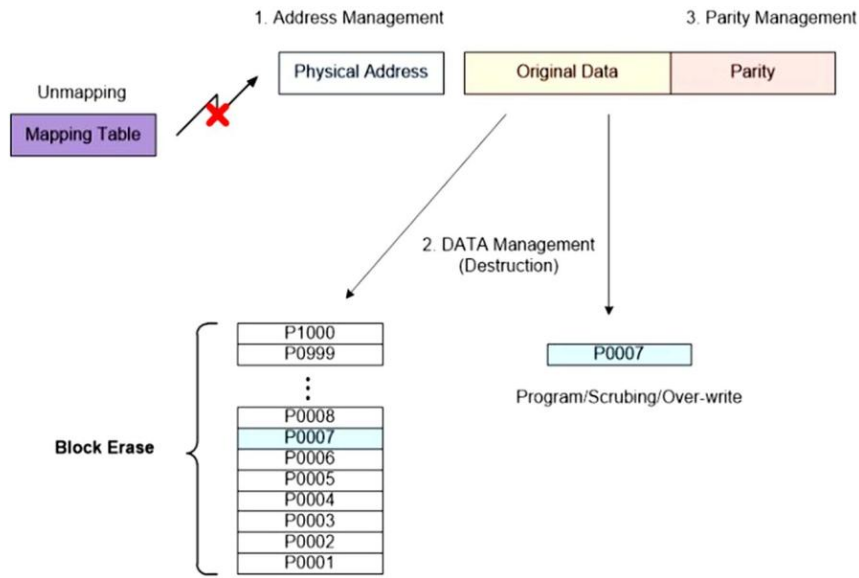

Figure 3: This presents a diagram distinguishing the various deletion techniques employed in storage devices.

Referring to Fig. 3, invalid data deletion techniques can primarily be categorized into Address Management, Data Management, and Parity Management techniques. Address Management techniques, while efficient in terms of management and performance, can potentially provide the lowest level of secure deletion capabilities due to prevalent data remnants within NAND flash memory. Data Management techniques focus on thoroughly deleting the original target data. Techniques such as block-erase operations, which delete at the block level, and partial programming/scrubbing/overwriting, which delete at the page level, can be employed. Block erase operations, which not only delete the invalid page (P0007) but also other surrounding pages, may entail a significant time delay before the respective block is entirely removed. Page-level deletion techniques, on the other hand, carry a high potential for causing program disturbance in neighboring valid pages. To reduce program disturbances in adjacent pages while enabling real-time secure deletion, Parity Management techniques are being introduced. These methods balance the trade-offs between security, efficiency, and reliability in modern storage systems.

The diagram illustrates a three-tier secure-deletion workflow—address management, data destruction, and parity management—operating in unison. First, the controller invalidates the logical-to-physical mapping entry, preventing any future read from reaching the target page. Second, the flash layer physically eradicates the content by either erasing the entire block containing the page or by repeatedly overwriting (scrubbing) the specific page to remove residual charge. Third, the system refreshes or invalidates parity/ECC metadata so that redundant information can no longer rebuild the deleted data while keeping the overall array consistent. Together, these steps make the content unreachable through mapping tables, unrecoverable in the cells, and absent from redundancy codes, thereby guaranteeing full and consistent sanitization.

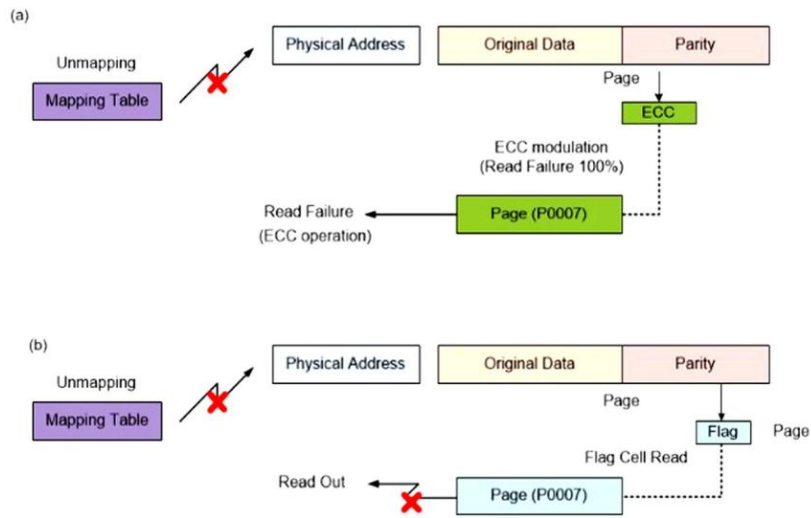

Figure 4: This conceptually illustrates the Parity Management techniques: (a) ECC Modulation method and (b) Flag Cell Read method.

**Fig. 4** illustrates two on-chip data sanitization approaches in NAND flash memory: (a) an ECC-based deletion technique using error induction, and (b) a flag-cell-based deletion technique using a dedicated indicator bit. These schemes logically erase data at the page level without immediately destroying the physical data, aiming to block any further read access to the target page. Both methods are designed to achieve real-time deletion at fine granularity, avoiding the performance and endurance costs of whole-block erase operations. (a) **ECC-based deletion** – The page (P0007) contains original data and parity (ECC) bits; by modulating the ECC code, any read to this page results in a 100% uncorrectable error (logical read failure). The physical address is unmapped from the file system, and the device's ECC logic treats the page as if it were irrecoverably corrupted (b) **Flag-cell-based deletion** – The page (P0007) likewise holds data and parity, but includes an extra *flag cell* in the spare area [15]. Setting this flag bit indicates the page is deleted; upon read, the controller detects the flag and marks the page as invalid, blocking any user read-out. In both cases, the logical mapping to the page is removed (FTL unmapped), preventing normal access.

In an ECC-based secure-deletion scheme, the flash controller deliberately corrupts a page's parity bits so its contents can no longer be decoded. Each NAND page holds user data and corresponding ECC, and the controller first unmaps the logical address before "poisoning" that parity. These bit flips push any later read beyond the code's correction limit, triggering an uncorrectable-error flag. The ECC decoder thus fails every read, making the page look garbled or empty even though the raw data still sits in the cells. Because only parity is altered, cell wear and disturb are minimal, and the controller can verify deletion instantly when the error appears. A determined attacker with direct flash access could still extract the underlying bytes, so the method is anti-forensic rather than cryptographically final. Readability cannot be restored without erasing the whole block and reprogramming fresh parity, effectively locking the page until a full erase.

The flag-cell deletion scheme assigns a dedicated metadata bit to each flash page that flips from 1 (valid) to 0 (deleted), marking the page as inaccessible without altering user data or ECC. When a delete request targets, say, page P0007, the controller first unmaps its logical address and then programs the page's flag to 0. Any later read checks this bit and, if it is 0, returns an error or zeros, blocking access while leaving the underlying cells intact. Hardware-wise, the flag steers a switch that isolates the page's data latch from the bus, and because the bit lives in spare area it is fetched in parallel with data, adding no extra latency. Once programmed, the flag cannot revert to 1 except via a full block erase, so the page remains locked until that erase occurs. This single-bit program is extremely fast, incurs no additional wear, and makes the page appear erased with negligible performance overhead. However, the raw bytes still exist, so vendors store redundant SLC copies of the flag and use majority voting to avoid spurious flips, yet highly sensitive data still needs stronger sanitization.

ECC-modulation and flag-cell techniques both let an SSD invalidate a single page on-chip without an immediate block erase. Each supports real-time TRIM-style deletion with negligible performance loss. ECC modulation "poisons" the page by corrupting its parity so the decoder can never reconstruct the data, whereas the flag cell flips a dedicated bit that blocks output. Because the flag method leaves data and parity intact, it preserves integrity better than ECC poisoning, which intentionally violates it. ECC poisoning can sometimes be deployed through firmware alone if partial spare-area writes are allowed, while flag cells demand dedicated hardware in the NAND and controller. Neither method clears the stored charges, so forensic attackers could still extract raw cells unless encryption or later garbage-collection erasure is added. Nevertheless, used separately or together, these fast, low-overhead schemes give flash systems practical secure-deletion options, each with its own trade-offs in complexity, security depth, and cost.

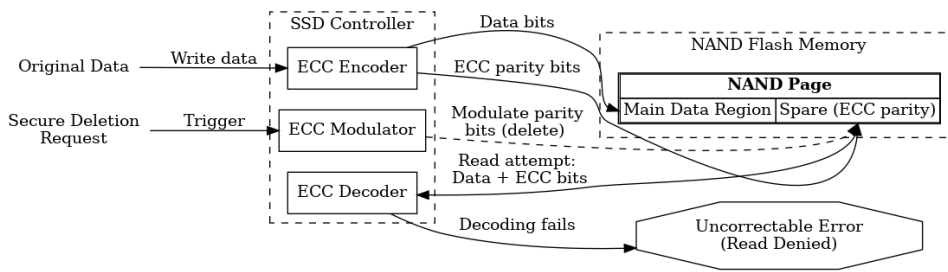

**Figure 5:** This conceptually illustrates ECC Modulation Secure Deletion Flow.

Referring to Fig. 5, this shows internal flow of the ECC modulation technique for secure deletion in an SSD. The SSD controller computes ECC parity for each NAND flash page and stores it in the page's spare area. To securely **delete** data, the controller's **ECC modulator** alters the stored parity bits (via a partial programming in the spare area) without erasing the page. As a result, any subsequent **read** of that page will fail ECC decoding (error correction becomes impossible), causing the device to return a **read-fail signal** to the host. This "read denial" effect serves as immediate confirmation that the original data is irrecoverable via normal means, achieving an anti-forensic deletion without the high latency of a block erase.

## 5 SECURE DELETION TECHNIQUES AND PERFORMANCE COMPARISON

Secure deletion techniques fall into three categories: address management (logical unmapping), data management (physical erasure or overwrite), and parity management (modifying ECC or redundancy metadata). ECC modulation deliberately corrupts parity to block reads, while flag-cell schemes mark pages invalid without altering the payload, offering real-time deletion with modest performance cost but differing security levels and hardware complexity. Our study compares these methods on effectiveness, speed, endurance impact, verifiability, and implementation cost to guide optimal selection. Rigorous evaluation must confirm complete data removal, quantify latency and wear, assess resource overhead, and factor in total financial and operational expenses.

Table 2: performance comparison of secure deletion techniques

| Target | | Efficacy | Speed | Durability | Overhead | Verification | Cost |
|---|---|---|---|---|---|---|---|
| Address Unmap | | 1 | 5 | 5 | 1 | 1 | 1 |
| Data | Encryption | 1 | 1 | 5 | 5 | 3 | 5 |
| | Erase | 5 | 1 | 1 | 1 | 5 | 5 |
| | Scrubbing | 3 | 5 | 3 | 3 | 1 | 3 |
| | Overwrite | 5 | 1 | 3 | 5 | 5 | 5 |
| | Deletion Pulse | 5 | 5 | 3 | 3 | 1 | 1 |
| Parity | ECC modulation | 5 | 5 | 5 | 3 | 5 | 3 |
| | Flag locking | 5 | 5 | 5 | 5 | 5 | 5 |

Table 2 provides a performance comparison of secure deletion techniques. The targets of secure deletion are broadly categorized into address, data, and parity. Secure deletion techniques for each target include address unmap, encryption, erase, scrubbing, overwrite, deletion pulse, ECC modulation, and block lock. For each secure deletion method, scores were assigned for efficiency, processing speed, durability, overhead, verifiability, and operational cost, rated as High, Medium, or Low.

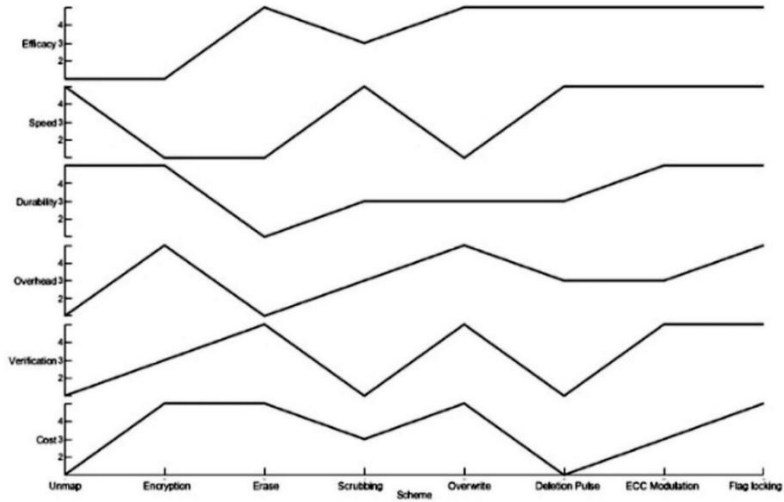

Figure 6: This presents a comparison of the performance for each secure deletion technique.

Figure 6 highlights the need to calibrate privacy levels against clear performance metrics, letting administrators tailor a storage device's security posture to real-world demands: for organizations handling sensitive data, high erase efficiency minimizes the chance of recovery after reuse or intrusion, and real-time workloads likewise demand instantaneous secure deletion; by contrast, consumer or non-commercial contexts may value cost, convenience, or energy use more than speed, while environments with relentless write–delete cycles hinge on durability, because heavy-overhead sanitization can throttle throughput; regardless of context, rigorous verification is essential to prove that a chosen technique delivers its promised protection, followed by cost–benefit analysis to keep solutions within budget and, where applicable, compliant with sectoral or national regulations; looking ahead, adaptive-privacy drives are poised to converge with homomorphic-encryption hardware, maintaining data in encrypted form even during computation and enabling safe analytics in privacy-sensitive domains such as healthcare, finance, cloud services, and big-data processing.

## 6 IMPLICATIONS AND FUTURE DIRECTIONS

Secure deletion, operational cost, and SSD longevity are in tension, making adaptive privacy strategies preferable to one-size-fits-all approaches. Our proposed framework delivers context-aware privacy control that adjusts to data sensitivity and workload characteristics, enhancing security without degrading throughput or write endurance. Empirical validation of various deletion mechanisms remains a priority for future work, and we aim to couple the framework with AI-driven privacy policies that learn usage patterns in real time. Another research avenue is the integration of homomorphic encryption so computations can occur directly on ciphertext, supporting privacy-preserving analytics for sensitive domains like healthcare and finance. Our analysis confirms that uniform deletion policies are suboptimal across the heterogeneous landscape of IoT storage; adaptive techniques instead strike a more efficient balance between privacy guarantees and system overhead. We plan to prototype an adaptive SSD and measure deletion efficacy, latency, and wear on real hardware, while incorporating key-bound encryption that can erase or revoke keys automatically when privacy levels change. Together, these advances will yield verifiable anti-forensic SSDs that inspire greater trust in next-generation IoT data management.

## 7 CONCLUSIONS

IoT devices now permeate everyday life. Consequently, securely deleting personal data on their storage media is paramount for privacy. This paper introduces an adaptive privacy-preserving SSD framework. The framework lets users flexibly select deletion techniques that match their privacy targets and cost constraints. We systematically compare address-, data-, and parity-management approaches. The results show that tailoring deletion methods to specific privacy requirements balances security with performance and endurance. Existing techniques struggle in IoT scenarios, but our adaptive strategy offers a scalable alternative. Future work will empirically test deletion efficacy, add AI-driven privacy policies, and extend the design to emerging storage technologies. We expect these efforts to hasten the arrival of robust anti-forensic SSDs and strengthen trust in next-generation IoT ecosystems.

## ACKNOWLEDGMENTS

We sincerely appreciate all researchers studying anti-forensics in relation to IoT.